# RR-DnCNN v2.0: Enhanced Restoration-Reconstruction Deep Neural Network for Down-Sampling-Based Video Coding

Man M. Ho, Jinjia Zhou, *Member, IEEE*, and Gang He

*Abstract*—Integrating deep learning techniques into the video coding framework gains significant improvement compared to the standard compression techniques, especially applying super-resolution (up-sampling) to down-sampling based video coding as post-processing. However, besides up-sampling degradation, the various artifacts brought from compression make super-resolution problem more difficult to solve. The straightforward solution is to integrate the artifact removal techniques before super-resolution. However, some helpful features may be removed together, degrading the super-resolution performance. To address this problem, we proposed an end-to-end restoration-reconstruction deep neural network (RR-DnCNN) using the degradation-aware technique, which entirely solves degradation from compression and sub-sampling. Besides, we proved that the compression degradation produced by Random Access configuration is rich enough to cover other degradation types, such as Low Delay P and All Intra, for training. Since the straightforward network RR-DnCNN with many layers as a chain has poor learning capability suffering from the gradient vanishing problem, we redesign the network architecture to let reconstruction leverages the captured features from restoration using up-sampling skip connections. Our novel architecture is called restoration-reconstruction u-shaped deep neural network (RR-DnCNN v2.0). As a result, our RR-DnCNN v2.0 outperforms the previous works and can attain 17.02% BD-rate reduction on UHD resolution for all-intra anchored by the standard H.265/HEVC. The source code is available at https://minhmanho.github.io/rrdncnn/.

## I. INTRODUCTION

VIDEO media has become one of the widest used applications in the digital era, depending on the development and popularization of video coding technology. Video coding technology has been iteratively developed for nearly 30 years and has continued the hybrid coding architecture

Manuscript received January 27, 2020; revised July 14, 2020, September 26, 2020, and December 9, 2020; accepted December 15, 2020. Date of publication January 8, 2021; date of current version January 14, 2021. This work was supported by the JST, PRESTO, under Grant JPMJPR1757 Japan. The associate editor coordinating the review of this manuscript and approving it for publication was Dr. Chaker Larabi. *(Corresponding author: Jinjia Zhou.)*

Man M. Ho and Jinjia Zhou are with the Graduate School of Science and Engineering, Hosei University, Koganei 184-8584, Japan (e-mail: jinjia.zhou.35@hosei.ac.jp).

Gang He is with the State Key Laboratory of Integrated Service Network, Xidian University, Xi'an 710071, China.

This article has supplementary downloadable material available at https://doi.org/10.1109/TIP.2020.3046872, provided by the authors.

Digital Object Identifier 10.1109/TIP.2020.3046872

of transform coding and predictive coding. Currently, video playback devices are increasingly diversified; however, network bandwidth and storage size are limited under many usage scenarios. Even with the popular advanced coding standard H.265/HEVC, the reconstructed video quality is still poor under extreme bandwidth conditions because it is sacrificed for a higher compression ratio. Therefore, a more efficient framework is required to reduce the bit-rate and maintain high video quality. There are two significant challenges: reducing the various distortions that result from video compression and increasing the compression ratio. With the development of deep learning techniques in recent years, CNN-based denoising and super-resolution have become suitable to address these challenges.

### A. Down-Sampling Based Coding (DBC)

Shen *et al.* [1] proposed the seminal Down-sampling Based Coding (DBC) framework, where a super-resolution technique is employed to restore the down-sampled frames to their original resolutions. Recently, deep learning-based super-resolution techniques outperform traditional methods and inspire researchers to improve the DBC framework. Li *et al.* [2] proposed CNN-based block up-sampling for intra-frame coding. Meanwhile, Lin *et al.* [3] improved [2] 's work by leveraging information between frames as block-level down- and up-sampling into inter-frame coding. Beyond leveraging information at block-level, Feng *et al.* [16] applied a frame-based DBC system with an extra enhancement network to remove compression artifacts before super-resolution. However, two separate networks are not linked well, and some useful features may be removed together with the artifacts degrading the super-resolution performance. We thus propose an end-to-end deep neural network to fully address compression degradation and enhance super-resolution.

### B. Deep Learning Approach for Reducing Compression Artifacts

Images/video resolution has rapidly increased from 480p and 720p to 1080p, 4K, and 8K. The frame rate has also increased from 30 fps to 60 fps and 120 fps. Under limited bandwidth, videos are encoded with a high compression ratio by sacrificing quality. According to recent CNN-based super-resolution achievements, transferring the low-size bitstream for high-resolution images/videos is possible. Similar

| DLR | Decoded Low-Resolution |
|-----|------------------------|
| LR | Low-Resolution |
| HR | High-Resolution |
| MSE | Mean Square Error |
| PSNR | Peak Signal-to-Noise Ratio |
| SSIM | The Structural Similarity Index |
| QP | Quantization Parameter |
| HEVC | High Efficiency Video Coding |
| HM | HEVC Test Model |
| UHD | Ultra High Definition |
| CNN | Convolutional Neural Network |
| CL | Convolutional Layer |
| RA | Random Access |
| LDP | Low Delay P |
| AI | All Intra |

to the related concept [16], we down-sample the source video before encoding, then restore, up-sample, and reconstruct images/videos after decoding. Thus, the bitstream capacity is much lower. Furthermore, the images/videos still meet quality requirements compared to the standard H.265/HEVC.

### C. Single Image Super-Resolution (SISR)

SISR recovers a high-resolution (HR) image based on a given low-resolution (LR) image. Being powered by deep learning, CNN-based SISR has generated surprisingly well-restored results. For example, Dong et al. [9] proposed a CNN-based SRCNN network structure to learn an end-to-end mapping from LR to HR and first proved the high-quality HR recovered using deep learning techniques. The network includes three layers: patch extraction, nonlinear mapping, and reconstruction. Afterward, Dong et al. [10] proposed a network named FSRCNN, which applies super-resolution in real-time. Kim et al. [11] presented a VDSR network, which can effectively improve image performance by learning residuals and increasing network depth to 20 layers; besides, they applied adjustable gradient clipping to solve their convergence problem. Zhang et al. [12] proposed the DnCNN network using residual learning outperforming VDSR in super-resolution. Kim et al. [13] proposed the deeply recursive convolutional network (DRCN), a very deep recursive layer via a chain structure with up to 16 recursions. Lai et al. [14] studied a Laplacian pyramid-based (LapSRN) image super-resolution network gradually up-sampling LR inside the networks. They thus provide low computational loads with high performance. Zhang et al. [15] adopted the attentional mechanism in deep learning to propose the very deep residual channel attention network (RCAN) exploiting abundant low-frequency information for super-resolution. However, these works mostly perform on typical degradation (e.g., bicubic, bi-linear) and forgo or naively train their models on other distortions, which usually appear in daily multimedia, such as noises, video compression artifacts, and JPEG compression. Consequently, the existing super-resolution works result in poor performance on unseen distortions. This work investigates the various degradation brought from video compression for training and testing a deep neural network.

### D. Recent Super-Resolution Works in Handling Degradation

To address the various types of degradation in super-resolution, Zhang et al. [5] synthesized bicubic degradation and Gaussian noise maps and fed them into the network to train together with LR. Zhao et al. [6] proposed an unsupervised learning network to learn unseen degradation and reconstruct the output. Bulat et al. [7] adopted the concept of generative adversarial networks (GANs) to learn how to degrade and down-sample high-resolution images and obtain a specific degradation for their super-resolution network. Meanwhile, Chen et al. [8] directly trained their models on JPEG degradation using an end-to-end deep convolutional neural network. Regarding the DBC field, besides up-sampling degradation, we also cope with various types of degradation, such as blocking artifacts and ringing artifacts, produced by video compression techniques. The most similar work [16] applied a refinement network before super-resolution to reduce compression artifacts; thus, the bicubic degradation of decoded images/videos is more precise. However, they still suffer from distortion due to imperfect refinement. To address this problem, we propose an end-to-end restoration-reconstruction deep neural network (RR-DnCNN) [18] using the degradation-aware technique as a two-loss function: restoration and reconstruction. In that, we treat the uncompressed LR as a transitional ground-truth inside our network. Our work is thus capable of effectively dealing with video compression distortion and up-sampling degradation.

### E. Skip Connections

Skip connections are widely used in modern deep network architectures. The well-known U-Net [21] and ResNet [22] are the seminal works proving the proficiency of skip (or shortcut) connections passing extracted features from shallower layers to higher layers. Consequently, in training, very deep neural networks can avoid vanishing gradients in back-propagation. Afterward, many variants are invented and successfully applied in many fields [15], [23]–[31], [34]. Notably, inspired by the attentional mechanism, the networks [15], [31] learn residual-attentional information to enhance their feature maps. The AvatarNet [33] leverages skip connections to stylize high-level features using shallow features in image style transfer. Iizuka et al. [34] fused well-trained features on image classification to consolidate the colorization task. Subjectively, Ho et al. [29] added hyperparameters for skip connections to visualize the shallow feature's effects on the stylized result, creating an adjustable image style transfer.

In this work, we enhance the learning capability of our prior work RR-DnCNN [18] by adopting the concept of skip connections [21]. Particularly, from the straightforward RR-DnCNN, we redesign the network architecture to have an u-shaped form and utilize up-sampling skip connections to pass the useful features captured by restoration to reconstruction. The novel network is called the restoration-reconstruction u-shaped deep neural network (RR-DnCNN v2.0).

### F. Contributions

The main contributions of this paper include the following aspects.

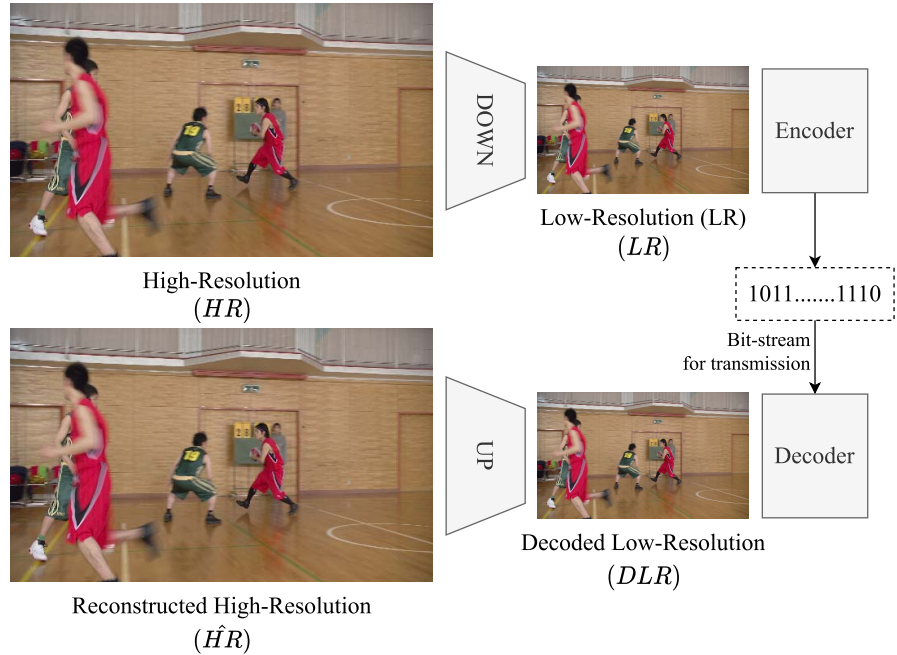

Fig. 1. The concept of down-sampling based coding. In our work, *DOWN* (bicubic interpolation) reduces compression complexity, "*Encoder*"/"*Decoder*" (HEVC 16.20) are for compression, and "*UP*" (our RR-DnCNN v2.0) compensates for the missing information caused by "*DOWN*" and compression.

- We proposed a degradation-aware technique [18] treating the original $LR$ as a transitional ground-truth to entirely solve the degradation from compression and up-sampling in down-sampling based video coding.
- On which compression configuration for training, we proved that the RA provides various degradation enough to cover other types.
- We improve our prior RR-DnCNN [18] to the degradation-aware restoration-reconstruction u-shaped deep neural network (RR-DnCNN v2.0) which leverages up-sampling skip connections to enhance the learning capability.
- Our novel down-sampling based coding outperforms previous works [2]–[4] and attains **17.02%** bit-rate reduction for all-intra at the low bit-rate range anchored by the standard H.265/HEVC.

## II. PROPOSED VIDEO CODING SYSTEM

### A. System Overview

In our study, we leverage the superiority of learned super-resolution techniques to reduce the bit-rate and enhance the video quality for our down-sampling based video coding system. Therefore, our video coding framework consists of down-sampling, HEVC codec, and a super-resolution network. We first use bicubic interpolation to down-sample high-resolution (HR) to have its low-resolution (LR), then compress the LR using HEVC codec. All sequences are down-sampled and compressed with their colors in YUV420 format. After decoding the bitstream, the super-resolution network removes compression artifacts and maps the decoded low-resolution (DLR) to its original HR at the decoding end, as shown in Figure 1. In previous work [18], we designed an end-to-end restoration-reconstruction deep neural network (RR-DnCNN) using the proposed degradation-aware technique to deal with compression degradation and up-sampling degradation entirely. However, a very deep neural network with many straightforward layers meets the gradient vanishing problem [22] in back-propagation. Training with the transitional ground-truth LR may solve the problem; however, shallow layers' useful features are absent in reconstruction. Consequently, restoration-reconstruction performance is limited and quickly saturated while training. To address the problem, inspired by U-Net [21], we utilize the up-sampling skip connections to pass the captured features from restoration into reconstruction for the learning capability robustness. Our novel network architecture is called the restoration-reconstruction u-shaped deep neural network (RR-DnCNN v2.0).

Unlike conventional super-resolution converting $DLR \rightarrow HR$ directly, our degradation-aware technique breaks the stage to $DLR \rightarrow LR \rightarrow HR$, where $LR$ is treated as a transitional ground-truth. As an advantage, our up-sampled low-resolution and features inside the network are enhanced for reconstruction. Our network is thus more robust than other works directly synthesizing $HR$ from $DLR$.

In YUV format, the luminance component is crucial for humans to see the objects in detail; therefore, we only train and test on the Y component. Initially, the original Y component denoted as $HR \in \mathbb{R}^{H \times W \times 1}$ is down-sampled at the scale of 2 using bicubic interpolation to have $LR \in \mathbb{R}^{H//2 \times W//2 \times 1}$, then encoded with the standard codec HEVC/H.265. After decoding, we feed $DLR \in \mathbb{R}^{H//2 \times W//2 \times 1}$ into our RR-DnCNN v2.0 denoted as $h$ to restore the missing information from compression, then generate the reconstructed $\hat{HR} \in \mathbb{R}^{H \times W \times 1}$. Our target is to minimize the error between $\hat{HR}$ and $HR$. Instead of entirely inferring from $DLR$ to $HR$, we present the degradation-aware technique treating the $LR$ as our transitional ground-truth. Additionally, residual learning is applied to speed up the network convergence, defined as:

$$\hat{LR}, R_{res}, R_{rec} = h(DLR) \qquad (1)$$

where $R_{res}$ represents the inferred residual between $LR$ and $DLR$ for restoration; meanwhile, $R_{rec}$ represents the inferred residual between up-sampled $\hat{LR}$ and $HR$. Inside our network, $DLR$ is restored to have $\hat{LR}$ as:

$$\hat{LR} = DLR + R_{res} \qquad (2)$$

then up-sampled by deconvolution and combined to reconstruction residual $R_{rec}$ to obtain the final $\hat{HR}$ as:

$$\hat{HR} = Deconvolution(\hat{LR}) + R_{rec} \qquad (3)$$

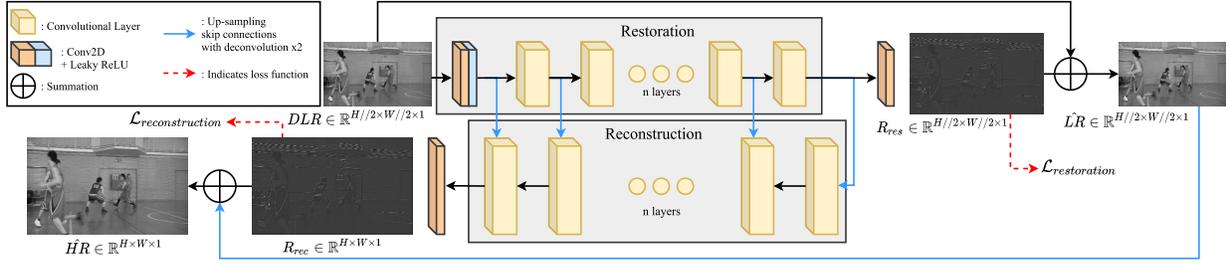

Fig. 2. Architecture of the proposed restoration-reconstruction u-shaped deep neural network (RR-DnCNN v2.0).

### B. Restoration-Reconstruction U-Shaped Deep Neural Network (RR-DnCNN v2.0)

Our previously proposed RR-DnCNN [18] treats uncompressed $LR$ as transitional ground-truth dividing network architecture into two parts: restoration and reconstruction. Particularly, restoration compensates for the lost information by video compression at low-resolution and then provides the feature-based information for reconstruction at high-resolution. However, the straightforward RR-DnCNN with many layers inferring one-way may face gradient vanishing in the back-propagation; plus, the useful features captured by the shallow layers of restoration are absent in reconstruction. To address the mentioned issues, inspired by U-Net [21], we leverage the up-sampling skip connections to pass captured features from restoration to reconstruction and improve RR-DnCNN to a restoration-reconstruction u-shaped deep neural network (RR-DnCNN v2.0). The new architecture RR-DnCNN v2.0 is thus more robust in learning capability than its old version.

In our RR-DnCNN v2.0 architecture, the restoration removes the compression artifacts from $DLR$ by learning the residual between $DLR$ and $LR$, which results in two directions: up-sampling features for reconstruction using deconvolutions and synthesizing a residual map to restore $DLR$ to have $\hat{LR}$. Subsequently, the reconstruction leverages up-sampled features from restoration to synthesize the residual between $HR$ and up-sampled $\hat{LR}$, then reconstruct up-sampled $\hat{LR}$ to have $\hat{HR}$, as illustrated in Figure 2.

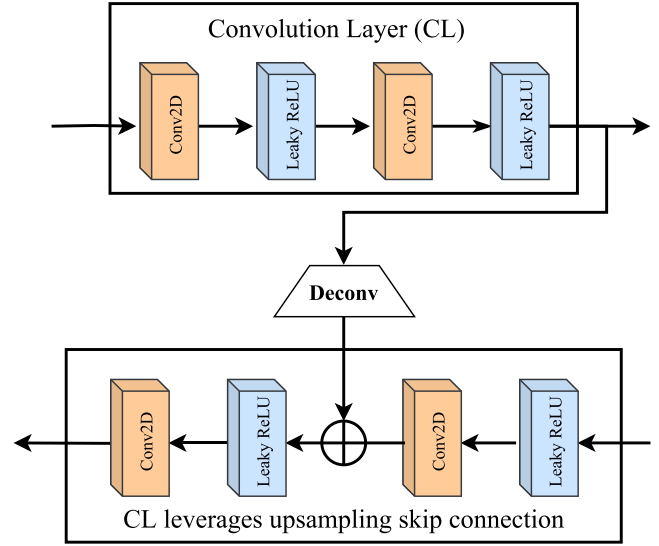

Fig. 3. Illustration of convolutional layers and up-sampling skip connections.

*1) Technical Details:* Our RR-DnCNN v2.0 consists of a convolution module and 10 convolution layers for repairing compressed video (restoration), and 10 convolution layers for super-resolution (reconstruction). Each convolution layer includes 2 convolution modules. Each convolution module is followed by Leaky Rectified Linear Units (Leaky ReLUs) with a negative slope of 0.01. Regarding transferring the captured features of each layer from restoration to reconstruction, we utilize up-sampling skip connections containing deconvolution with a kernel size of $3 \times 3$, a stride of 2, padding of 1, and out padding of 1. Each layer of reconstruction receives the captured features at its middle by summing, as illustrated in Figure 3. Unlike other convolution modules using a kernel size of $3 \times 3$, a stride of 1, and a padding of 1, the first convolution of restoration uses the kernel size of $5 \times 5$ and a padding of 2 to initially extract useful information with a higher receptive field. The depth channel of 64 is maintained throughout the network, as described in Figure 2.

### C. Loss Function

As described in previous work [18], our losses are computed using the mean square error as:

$$MSE = \frac{1}{N} \sum_{i=1}^{N} \| R_i - R_i^* \|_2^2 \qquad (4)$$

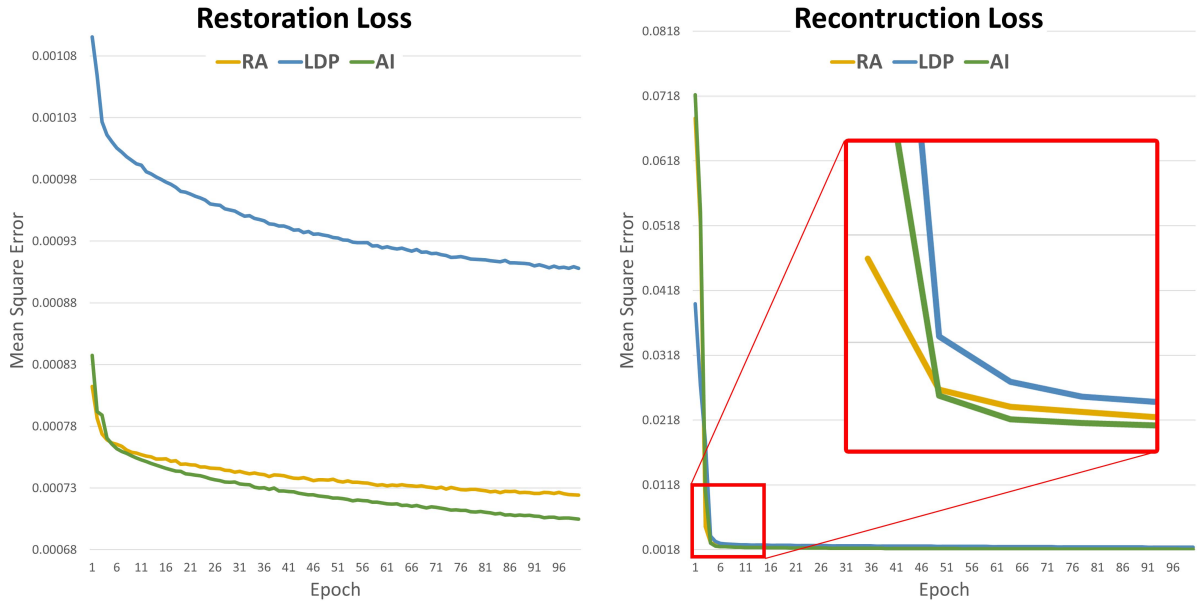

Fig. 4. Training processes of 3 domains RA, LDP, and AI at QP = 37 in mean square error. *Left*: restoration loss, *right*: reconstruction loss.

where N is the number of elements in a batch, $R_i$ represents the ground-truth residuals; meanwhile $R_i^*$ can be the $i^{th}$ $R_{res}$ or $i^{th}$ $R_{rec}$. Since our degradation-aware technique treats $LR$ as a transitional ground-truth in the middle of the network, we add loss weights of $\alpha$ and $\beta$ to balance optimizing errors between restoration and reconstruction. The total loss function is defined as:

$$\mathfrak{L} = \alpha * \mathfrak{L}_{\text{restoration}} + \beta * \mathfrak{L}_{\text{reconstruction}} \tag{5}$$

where $\mathfrak{L}_{\text{restoration}}$ minimizes the error between $(LR - DLR)$ and $R_{res}$, while $\mathfrak{L}_{\text{reconstruction}}$ minimizes the error between $(HR - Deconvolution(LR))$ and $R_{rec}$. Different from RR-DnCNN [18], RR-DnCNN v2.0 shows the faster convergence and more stable learning; therefore, we empirically set the weights in equation 5 as $\alpha = 0.5, \beta = 0.05$.

### D. Training Configuration

In previous work [18], we explored the characteristic of compression degradations from three default configurations **RA**, **LDP**, and **AI** compensating for the missing information of the current frame by leveraging **neighboring frames**, **previous frames**, and **neighboring pixels within the current frame**, respectively. As shown in Figure 4, training on sequences compressed by AI outperforms others in convergence; in contrast, sequences compressed by LDP are the most difficult to be learned. In the validation phase on *BasketballDrive*, the model trained on LDP works well on only its domain. Meanwhile, the model trained on RA has the best performance on both RA and AI, and outperforms others on average quantitatively, as shown in Table II. We conclude that the degradation information brought from RA is rich enough to generalize other degradation cases. Therefore, RA is utilized for training our models.

### TABLE II

Ablation Studies on Compression Configurations. We Train Our Models Using Configuration RA, LDP, and AI, and Test on Decoded Low-Resolution Video *BasketballDrive* 1920 × 1152 (Resized by Bicubic), Which Is Compressed in RA, LDP, and AI. The **bold**/<u>underline</u> Values Show the **best**/<u>worst</u> PSNR on Each Configuration. The Model Trained on RA Achieves the Best Performance on Average

| Training on | Test on | | | |
|---|---|---|---|---|
| | RA | LDP | AI | Average |
| RA | **31.99** | 31.83 | **33.21** | **32.34** |
| LDP | <u>31.92</u> | **31.84** | <u>33.06</u> | <u>32.27</u> |
| AI | 31.95 | <u>31.82</u> | 33.2 | 32.32 |

### III. Experiments, Comparison, and Results

#### A. Experiments

*1) Data Preparation:* Uncompressed sequences are crucial for providing reliable analysis and understanding of video compression degradation in learning. Therefore, we only select uncompressed sequences for training and testing. Our previous work RR-DnCNN [18] is trained on small-scale data with the small resolution CIF 352 × 288, which limits the performance on the larger resolution. Moreover, compressing small-size videos at a low bit-rate gives very large distortion leading to weak learning capability. Therefore, this work has two stages of training:

*a) First stage: Training on large-scale uncompressed videos in CIF:* Beside 34 uncompressed videos as 18,478 frames from Xiph Video Test Media in CIF 352 × 288 used in [18], we add 1,912 resized frames from class D, including the sequences *BlowingBubbles*, *RaceHorses*, *BQSquare*, and *BasketballPass*. Regarding down-sampling, the size of *H R* 352 × 288 is down-sampled at scale × 2 to obtain the size of *L R* 176 × 144 using bicubic interpolation.

*b) Second stage: Fine-tuning on uncompressed videos at a larger resolution:* We fine-tune our best model from the first stage on 11 uncompressed videos containing 3,300 frames from the SJTU sequences in UHD [20] (excluding the test sequences we used in Table VI). The training sequences are down-scaled from $3840 \times 2160$ to $1920 \times 1152$ for $HR$ and $960 \times 576$ for $LR$ using bicubic interpolation. Since we resize the UHD to HD, the distortion is not significant, and the resized sequences are still high-quality, like uncompressed ones.

For the evaluation, in addition to the test sequences from classes A, B, C, and E, we also conduct comparisons on UHD sequences from [20], such as *Campfirearty*, *Fountains*, *Runners*, *RushHour*, and *TrafficFlow*, and on additional class B sequences, such as *BlueSky*, *Pedestrian*, and *RushHour*, which the previous works [2]–[4], [18] did not consider.

Our work is based on the HEVC Test Model (HM) version 16.20. Since our scheme tends to encode the $\times 2$ down-sampled sequences, we crop the test sequences having the resolution of $1920 \times 1080$ to $1920 \times 1072$ to satisfy the coding unit (CU) requirement. All experiments are conducted on only the Y component, being crucial for human eyes in seeing details. Besides, the color channels U and V are up-sampled from $DLR$ using the nearest-neighbor interpolation to illustrate reconstructed videos with color.

*2) Data Augmentation:* To augment training data, we apply random crop $120 \times 120$ for training our models on CIF resolution (first stage), and $512 \times 512$ for fine-tuning our pre-trained models on UHD sequences (second stage). Afterward, we utilize random flip in horizontal and vertical ways and random rotation in 0, 90, 180, and 270 degrees. Finally, Y values are normalized in the range [0, 1].

*3) Training Details:* We change the Adam optimizer [17] previously used in [18] to the Rectified Adam optimizer (RAdam) [19] with learning warm-up, which helps to avoid convergence problem at early stage of training, with an initial learning rate of 0.0001, coefficients $\beta_1 = 0.9, \beta_2 = 0.999$. Besides, we use the batch size of 16 correspondings to 1275 iterations per epoch at the first stage of training on CIF sequences. Meanwhile, at the second stage of fine-tuning on a larger resolution, the batch size is set to 2 respective to 1477 iterations per epoch. Every 100 epochs require approximately 43 hours on a Tesla V100 GPU.

### B. Comparison Between Our Novel Network and Its Old Version RR-DnCNN [18] in Learning Capability

Based on our previous work [18], we conduct ablation studies on training with/without RAdam [19], our novel network architecture RR-DnCNN v2.0 with up-sampling skip connections, and 2-stage training/fine-tuning. Besides, we train ablation models under the same condition. Particularly, we compress sequences using RA configuration at QP = 37, which is considered to have enough useful information to cover other QPs, as proved in Section II-D.

*1) On the Changed Optimizer:* We change the optimizer from Adam [17] to Rectified Adam (RAdam) [19], a variant of Adam, with learning rate warm-up stage. As a result, our

### TABLE III

Ablation Study and Comparison Between Bicubic, DnCNN [12], and Our Works in PSNR on *BasketBallDrive* $1920 \times 1152$ (up-Sampled by bicubic) Sequence Compressed at QP = 37 Using RA, LDP, and AI. Our Network (RR-DnCNN v2.0 With up-Sampling Skip Connections) Outperforms Others as Shown With **bold** Values

| Method | PSNR (dB) | | | |
|---|---|---|---|---|
| | RA | LDP | AI | Avg. |
| Bicubic | 31.48 | 31.37 | 32.63 | 31.83 |
| DnCNN [13] | 31.6 | 31.46 | 32.64 | 31.9 |
| RR-DnCNN [20] | 31.99 | 31.83 | 33.21 | 32.34 |
| RR-DnCNN [20] + RAdam [21] | 32.08 | 31.88 | 33.28 | 32.41 |
| RR-DnCNN v2.0 + RAdam [21] | 32.12 | 31.94 | 33.36 | 32.47 |
| RR-DnCNN v2.0 + RAdam [21] + 2-stage training/fine-tuning | **32.25** | **32.07** | **33.50** | **32.6** |

model can be converged more effectively with higher performance compared to [18], as shown in Table III (*rows 4,5*).

*2) On the Novel Network Architecture:* Although our previous work RR-DnCNN [18] outperforms the HEVC, its network architecture still has a limitation on learning capacity due to straightforward design with many layers, which easily meets gradient vanishing. Plus, essential features that restoration has been captured in its first shallow layers may be absent in reconstruction. Therefore, inspired by U-Net [21], we design specific network architecture using up-sampling skip connections, called restoration-reconstruction u-shaped deep neural network (RR-DnCNN v2.0), to enhance the learning capability, as described in Section II-B. To prove the effectiveness of our novel architecture, we train RR-DnCNN and RR-DnCNN v2.0 under the same conditions and validate them on sequences in $1920 \times 1152$ such as *BlueSky*, *Pedestrian*, *RushHour*, *BQTerrace*, *BasketballDrive*, *Cactus*, *Kimono*, *ParkScene* resized by bicubic interpolation. As a result, our RR-DnCNN v2.0 shows better convergence in restoration and reconstruction. Furthermore, the validation error of RR-DnCNN v2.0 is lower than its old version on average, as shown in Figure 5. In conclusion, our novel network RR-DnCNN v2.0 outperforms RR-DnCNN [18] in learning capability.

*3) On 2-Stage Training / Fine-Tuning:* Down-sampling a small-size sequence causes more lossy information than down-sampling a large-size one. Besides, after compression, details are more unrecognizable, even by humans. Therefore, our performance is limited when training on a small resolution of $352 \times 288$. To address the problem, we collect more uncompressed sequences from SJTU UHD [20] in $3840 \times 2040$ excluding the compared sequences in Table VI, then resize them to HD $1920 \times 1152$. We thus have more large-size high-quality videos for fine-tuning the best model from the first stage. As a result, the fine-tuned model of RR-DnCNN v2.0 outperforms all designed models quantitatively with the highest PSNR as **32.6** dB, as shown in Table III.

### C. Results

In the standard video codec, a specific frame of a video is compressed and restored in three ways: 1) leveraging neighboring frames (RA), 2) previous frames (LDP), and

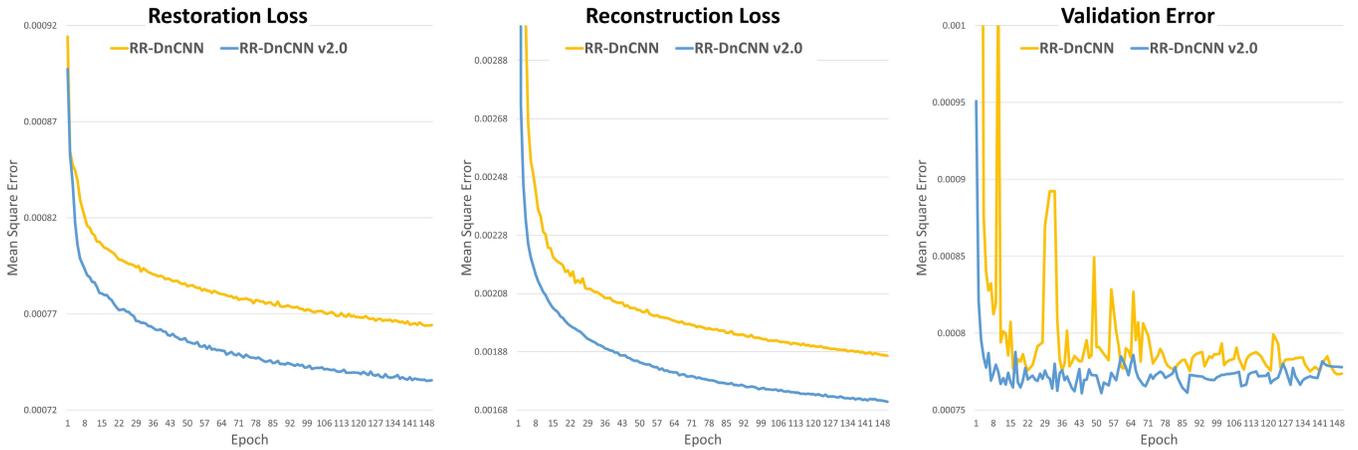

Fig. 5. Ablation study on learning capability of the network architecture compared to RR-DnCNN [18]. We train two networks in the same condition and visualize its restoration loss (left), reconstruction loss (middle), and validation error (right). Our novel RR-DnCNN v2.0 outperforms its old version in learning capability.

3) pixels within that frame (AI), as three standard configurations in video codec. Each configuration provides a characteristic degradation distorting frame structure such as blocking, ringing artifacts. To prove our proficiency in solving all mentioned types of degradation, we evaluate the ablation models using PSNR and previous works on sequences compressed by three configurations RA, LDP, and AI. Particularly, we compare our works to DnCNN [12] and our previous RR-DnCNN [18], which are our baselines, and to the previous works [2]–[4], [18] in compression proficiency BD-rate anchored by the standard H.265/HEVC. Since the previous works use a different version of the HEVC Test Model (HM) software from ours, we also experiment on both HM versions to make our comparison fair. Furthermore, we qualitatively compare to the standard H.265/HEVC by showing reconstructed videos in the approximate bit-rate. Note that we only train and test on the Y component; therefore, the similarity scores and BD-rate are calculated based on the Y channel. Besides, we up-sample U, V components of DLR using the nearest-neighbor interpolation for our color information in qualitative comparisons throughout this work.

*1) Compared to the Baseline Architecture DnCNN [12] and Our Previous Work RR-DnCNN [18]:* Previously in the work [18], we have proved the outperformance of RR-DnCNN's network architecture compared to DnCNN [12] trained in the same condition, as re-detailed in Table III (*rows 1, 2, 3 for methods*). Although the previous work RR-DnCNN [18] shows its proficiency in reducing artifacts and super-resolution, its straightforward deep network architecture having too many layers inferring in one-direction causes gradient vanishing; plus, the useful features captured by the shallow layers are absent in reconstruction. Consequently, the performance is easily saturated. To enhance the learning capability, we utilize up-sampling skip connections to pass the feature maps from restoration for reconstruction. Our RR-DnCNN v2.0 thus shows the greater effectiveness in learning capability. As described in Table III, our RR-DnCNN v2.0 quantitatively outperforms others on three con-figurations. Furthermore, we also compare the works on two test sequences *PeopleOnTheStreet* and *Traffic* with resolution $2160 \times 1600$ both objectively (PSNR/SSIM) and subjectively. As an experimental result, our RR-DnCNN v2.0 shows clearer edges, shapes such as *folds and pattern on the shirt*, and *the text and lines*, compared to bicubic interpolation and RR-DnCNN [18], as illustrated in Figure 6. Regarding restoration at low-resolution, the RR-DnCNN v2.0 stably outperforms its previous version [18] on all sequences from class B and additional B. Moreover, we show a comparison in computational complexity, described in Appendix V.

*2) Compared to Related Works on Down-Sampling Based Video Coding:* To the best of our knowledge in down-sampling based video coding, the latest works [2]–[4] are similar to our work improving the standard HEVC codec in a low bit-rate range, especially when QPs = {32, 37, 42, 47}. Therefore, we quantitatively compare our work to [3], [18] on RA, LDP and [2], [4], [18] on AI configuration in bit-rate savings (BD-rate), anchored by the standard HEVC. The experimental results of related works are provided in their materials [2]–[4]. Since we use the HEVC test model version 16.20, which is different from previous works using version 12.1, we additionally experiment on both versions to prove that our results on HM 16.20 are comparable.

*a) Our result on HEVC test model (HM) versions 16.20 and 12.1:* The experimental results of previous works [2]–[4] in this paper are borrowed from their materials. However, our HM version is different. We thus conduct a comparison between both versions in BD-rate with QPs = {32,37,42,47} on the video *ParkScene*. As a condition, our results on HM 16.20 are comparable to the previous works' using HM 12.1 only if our results on HM 12.1 are similar or better than on HM 16.20. As a result, our performance on HM 12.1 is even better than on our used version 16.20 achieving 10.74% bit-rate reduction, as shown in Table IV. Therefore, the performance gap between the two HM versions does not prioritize our work, and our current results are comparable to the previous works.

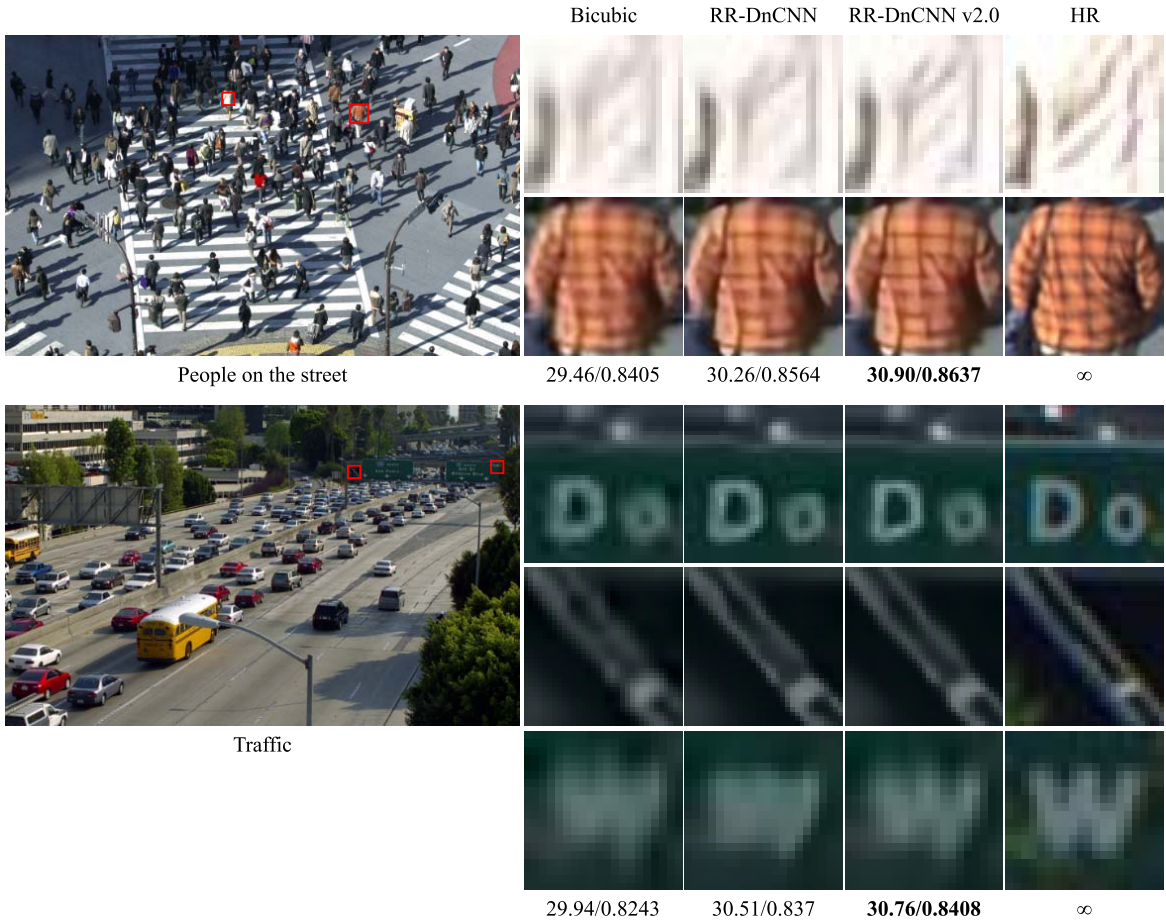

Fig. 6. Subjective comparison between bicubic, the previous work RR-DnCNN [18], and our RR-DnCNN v2.0 in PSNR (dB)/SSIM. The **bold** values represent the best performance.

TABLE IV
WE CONDUCT COMPARISON BETWEEN TWO HEVC TEST MODEL (HM) VERSIONS 16.20 AND 12.1 IN BD-RATE FOR QPs={32,37,42,47} ON THE VIDEO *ParkScene* TO ENSURE THAT THE PERFORMANCE GAP WILL NOT PRIORITIZE OUR RESULTS COMPARED TO PREVIOUS WORKS. AS A RESULT, OUR PERFORMANCE ON HM 12.1 IS EVEN BETTER THAN HM 16.20

| Seq. | QP | HM 16.20 | | | | | HM 12.1 | | | | |
|---|---|---|---|---|---|---|---|---|---|---|---|
| | | HEVC | | Ours | | BD-rate (%) | HEVC | | Ours | | BD-rate (%) |
| | | Bit-rate (kbps) | PSNR (dB) | Bit-rate (kbps) | PSNR (dB) | | Bit-rate (kbps) | PSNR (dB) | Bit-rate (kbps) | PSNR (dB) | |
| ParkScene 1920 × 1152 (bicubic) | 32 | 867.52 | 35.49 | 296.39 | 32.02 | -10.39 | 915.25 | 35.29 | 308.97 | 31.81 | -10.74 |
| | 37 | 404.18 | 32.96 | 135.06 | 30.04 | | 429.20 | 32.73 | 138.96 | 29.79 | |
| | 42 | 186.36 | 30.60 | 59.10 | 28.10 | | 191.74 | 30.30 | 58.38 | 27.81 | |
| | 47 | 92.30 | 28.67 | 27.45 | 26.49 | | 76.95 | 28.10 | 22.50 | 26.07 | |

*b) Compared to [3], [18] on RA, LDP in BD-rate:* Given a current frame in a specific video, [3] proposed a block-based multiple-image super-resolution (MISR) technique having the advantage of leveraging neighboring frames to compensate for the current frame. However, they ignore compression degradation and roughly reconstruct decoded frames. It remains distorted by compression. Meanwhile, we propose a degradation-aware technique treating LR as a transitional ground-truth to restore DLR before up-sampling inside the network. Therefore, degradations from compression and up-sampling are distinguished and solved effectively. Besides, we improve the network architecture, change the optimizer to RAdam [19], and have 2-stage training/ fine-tuning. As a result, our RR-DnCNN v2.0 achieves better compression proficiency than its previous version RR-DnCNN [18] and block-based MISR [3] as BD-rates **−14.11%, −8.86%, −18.12%, −6.14%** on class **A, B, additional B**, and **average** respectively for RA. Furthermore, RR-DnCNN v2.0 outperforms the previous works with better BD-rates as **−13.01%, −10.06%, −17.37%, −5.58%, −7.43%** on class **A, B, additional B, C**, and **average** correspondingly for LDP, as shown in Table V.

*c) Compared [2], [4], [18] on AI in BD-rate:* We compare our works to the recent SISR-based works [2], [4] on sequences compressed using AI configuration. Unlike the compared works training on and solving degradation resulting from only AI, we train our RR-DnCNN [18] and its version 2.0 on sequences compressed using RA

TABLE V

Objective Comparison Between Our Work, RR-DnCNN [18], and the Work [3] on RA, LDP in BD-Rate Anchored by the Standard Video Coding HEVC. We Calculate the Average BD-Rate (%) for Classes A, B, Additional B, C, E. The Average BD-Rate (%) of Intersection Is Estimated on Classes A,B,C for RA, and Classes B, C, E for LDP. **Bold** Values Show the Best BD-Rate (%) on Average

| | Sequence | RA | | | LDP | | |
|---|---|---|---|---|---|---|---|
| | | [3] | RR-DnCNN [20] | RR-DnCNN v2.0 | [3] | RR-DnCNN [20] | RR-DnCNN v2.0 |
| Class A 2560 × 1600 | People | -6.60 | -4.73 | -13.49 | ∼ | -3.65 | -12.92 |
| | Traffic | -3.60 | -8.23 | -14.72 | ∼ | -7.53 | -13.09 |
| Class B 1920 × 1072 (cropped) | Kimono | -5.80 | -11.63 | -14.42 | -4.20 | -13.29 | -15.03 |
| | ParkScene | -2.30 | -7.36 | -12.86 | -2.30 | -9.17 | -12.86 |
| | Cactus | -3.80 | -2.53 | -7.78 | -3.80 | -2.09 | -6.79 |
| | BasketballDrive | -7.50 | 3.31 | -0.38 | -9.50 | -2.88 | -5.55 |
| Additional B 1920 × 1072 (cropped) | Bluesky | ∼ | -12.11 | -21.58 | ∼ | -12.30 | -20.83 |
| | Pedestrian | ∼ | -11.42 | -16.05 | ∼ | -11.25 | -16.41 |
| | RushHour | ∼ | -13.12 | -16.73 | ∼ | -10.94 | -14.88 |
| Class C 832 × 480 | BasketballDrill | -6.40 | -5.36 | -6.10 | -5.70 | -6.95 | -8.53 |
| | BQMall | -2.30 | 3.42 | 1.02 | -1.90 | -3.12 | -5.04 |
| | PartyScene | -2.30 | 17.73 | 18.03 | -2.20 | 2.98 | 2.75 |
| | RaceHorses | -5.90 | -11.22 | -10.72 | -2.90 | -11.84 | -11.51 |
| Class E 1280 × 720 | FourPeople | ∼ | 3.94 | -5.13 | -2.20 | -0.86 | -6.88 |
| | Johnny | ∼ | -7.42 | -11.93 | -4.00 | -12.93 | -15.81 |
| | KristenAndSara | ∼ | -0.97 | 13.42 | -2.20 | -6.62 | 3.53 |
| Avg. on class A | | -5.10 | -6.48 | **-14.11** | ∼ | -5.59 | **-13.01** |
| Avg. on class B | | -4.85 | -4.55 | **-8.86** | -4.95 | -6.86 | **-10.06** |
| Avg. on additional B | | ∼ | -12.21 | **-18.12** | ∼ | -11.49 | **-17.37** |
| Avg. on class C | | **-4.23** | 1.14 | 0.56 | -3.18 | -4.73 | **-5.58** |
| Avg. on class E | | ∼ | **-1.48** | -1.21 | -2.80 | **-6.81** | -6.39 |
| Avg. on Intersection | | -4.65 | -2.66 | **-6.14** | -3.72 | -6.07 | **-7.43** |

TABLE VI

Quantitative Comparison Between Our Work, the Work [2], [4] for All-Intra (AI). Our Method Outperforms Others and Attains **17.02%** BD-Rate (%) Reduction on UHD Resolution Compared to the Standard HEVC. **Bold** Values Show the Best BD-Rate (%) on Average

| | | [2] | [4] | RR-DnCNN [20] | RR-DnCNN v2.0 |
|---|---|---|---|---|---|
| Class A 2560 × 1600 | People | -9.70 | -9.50 | -5.60 | -16.30 |
| | Traffic | -10.10 | -12.40 | -9.72 | -15.19 |
| Class B 1920 × 1072 (cropped) | Kimono | -7.70 | -13.00 | -8.60 | -11.32 |
| | ParkScene | -7.10 | -8.80 | -6.79 | -11.69 |
| | Cactus | -6.60 | -7.10 | -5.52 | -9.86 |
| | BasketballDrive | -6.10 | 7.00 | 2.29 | -0.39 |
| Additional B 1920 × 1072 (cropped) | Bluesky | ∼ | ∼ | -12.30 | -20.83 |
| | Pedestrian | ∼ | ∼ | -11.25 | -16.41 |
| | RushHour | ∼ | ∼ | -10.94 | -14.88 |
| Class C 832 × 480 | BasketballDrill | -4.90 | -10.70 | -10.92 | -12.46 |
| | BQMall | -2.90 | 17.20 | -0.85 | -2.64 |
| | PartyScene | -1.00 | 8.90 | -1.78 | -0.88 |
| | RaceHorses | -6.70 | -6.80 | -10.46 | -8.52 |
| UHD [22] 3840 × 2160 | Campfire_Party | -8.40 | -25.40 | -26.64 | -29.10 |
| | Fountains | -4.00 | -5.40 | -4.08 | -8.78 |
| | Runners | -11.20 | -13.20 | -10.87 | -13.96 |
| | Rush_Hour | -8.50 | -14.60 | -11.18 | -14.15 |
| | Traffic_Flow | -12.70 | -16.00 | -14.34 | -19.11 |
| Class E 1280 × 720 | FourPeople | -7.20 | -3.90 | -2.71 | -8.98 |
| | Johnny | -9.00 | -8.10 | -9.63 | -12.53 |
| | KristenAndSara | -6.80 | -0.70 | -5.66 | 3.81 |
| Avg. on class A | | -9.90 | -10.95 | -7.66 | **-15.74** |
| Avg. on class B | | -6.88 | -5.48 | -4.65 | **-8.32** |
| Avg. on additional B | | ∼ | ∼ | -11.49 | **-17.37** |
| Avg. on class C | | -3.88 | 2.15 | -6.00 | **-6.13** |
| Avg. on class E | | **-7.67** | -4.23 | -6.00 | -5.90 |
| Avg. on UHD | | -8.96 | -14.92 | -13.42 | **-17.02** |
| Avg. on Intersection | | -7.26 | -6.81 | -7.95 | **-10.67** |

configuration to solve all types of degradation. As a result, our RR-DnCNN v2.0 outperforms its previous version [18] and the previous works [2], [4] with better BD-rates as −15.74%, −8.32%, −17.37%, −6.13%% on class **A, B, additional B, C** respectively. Furthermore, this work shows the most proficiency on **UHD** sequences and attains the average BD-rate −17.02% for AI, anchored by the standard

HEVC. In summary, our RR-DnCNN v2.0 outperforms the previous works as BD-rate **10.67%** on average, as shown in Table VI.

*3) Compared to the Standard HEVC:* In the previous Section III-C.2, our method outperforms the previous works, as well as the standard HEVC, quantitatively with better BD-rates as **−6.14%, −7.43%, −10.67%** on RA, LDP, and AI, respectively, and attains **17.02%** bit-rate reduction on the UHD for AI. For the easier measurement, we visualize the rate-distortion curves (R-D curves) of bicubic, RR-DnCNN [18], this work RR-DnCNN v2.0, and HEVC on *PeopleOnTheStreet* (People), *BlueSky*, *Kimono*, *CampfireParty*, and *RushHour* compressed by the configurations LDP and AI. As shown in Figure 7, down-sampling based video coding outperforms HEVC in compressing videos at low bit-rate, even with the efficient bicubic, as visualized on *BlueSky*, *Kimono*, *RushHour*, especially large-size video *Campfire Party*. However, the outperformance of bicubic is not significant and worse than HEVC in the video *PeopleOnTheStreet*. Although our previous work RR-DnCNN [18] can improve the performance, it still has a limitation on a sequence such as *PeopleOnTheStreet* compressed by AI. Meanwhile, the R-D curves of RR-DnCNN v2.0 show the best performance on all sequences.

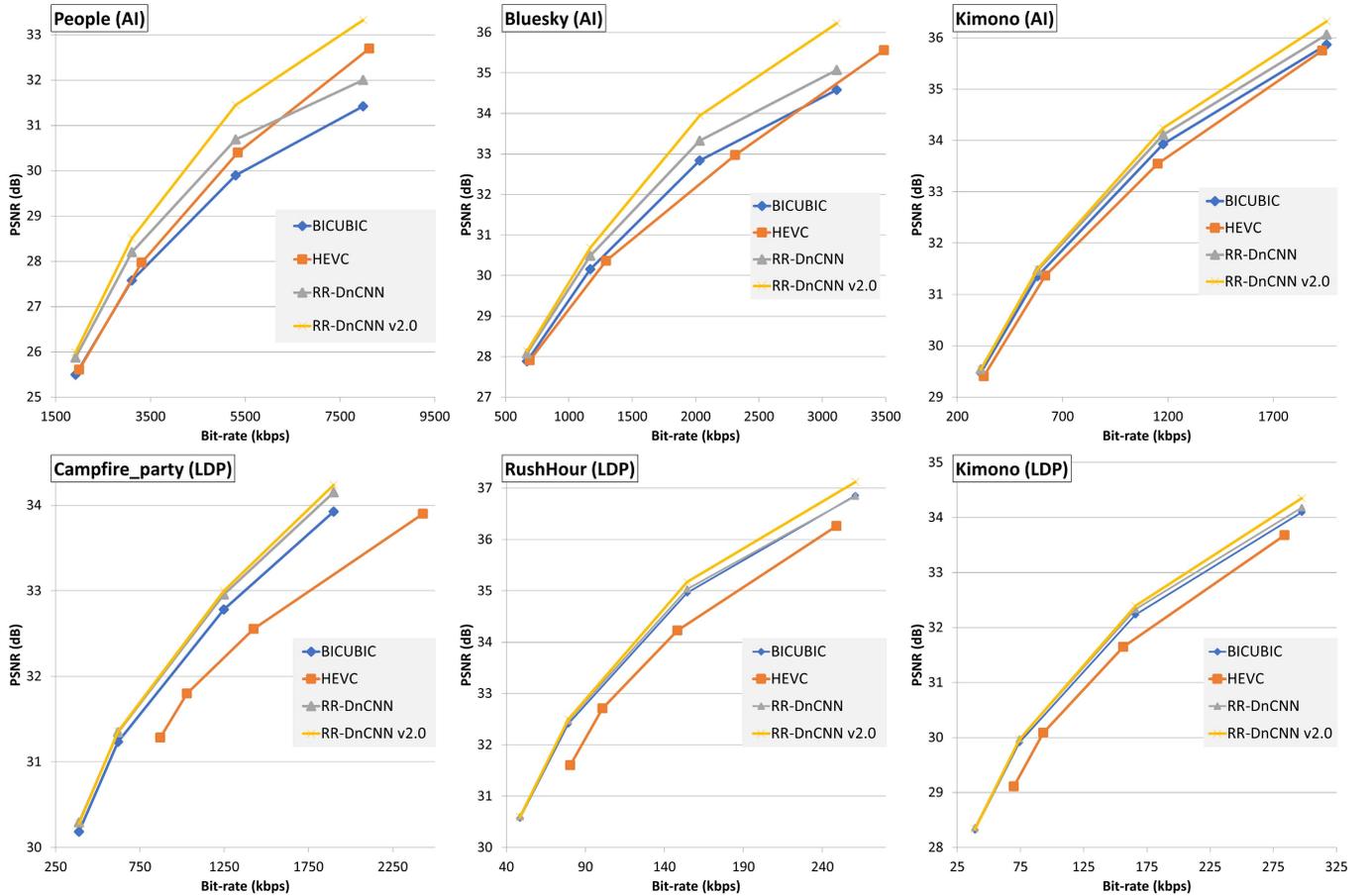

Fig. 7. Rate-distortion (R-D) curves of bicubic, RR-DnCNN [18], this work RR-DnCNN v2.0, and the standard HEVC on test sequences *People on the street*, *BlueSky*, *Kimono*, *Campfire party*, *RushHour*. Since compressing high-resolution (HEVC) provides much larger bit-rate than compressing low-resolution (bicubic, RR-DnCNN [18], and RR-DnCNN v2.0), for showing the performance gap between curves intuitively, we use QPs = {39,43,47,51} to draw HEVC's R-D curves, and QPs = {32,36,41,45} for other methods' curves. As shown, our method apparently outperforms HEVC. For each chart, the **x-axis** (horizontal) represents bit-rate (kbps); meanwhile, **y-axis** (vertical) represents PSNR (dB).

Our target is to outperform the standard HEVC in a low bit-rate range such as QPs = {32, 37, 42, 47}, which has degradation large enough for our method to exploit. Therefore, we consider using a QP of 37 to train our model for the best performance on all QPs. Although the well-trained model outperforms HEVC at low bit-rates, our performance, however, is reduced gradually, then worse than the standard HEVC when decreasing QP, as shown in our Appendix V-B. A possible way to address this problem is to train the model on mixed QPs.

Besides, we conduct a subjective comparison between our method and the standard HEVC in an approximate bit-rate condition on *Kimono*, *Pedestrian*, and *RushHour* compressed by RA, LDP, AI respectively. All videos are resized using bicubic interpolation to have the size of 1920 × 1152. Additionally, we measure bit-rate (kbps)/PSNR (dB)/SSIM for each method in every compared video. As a result, our results show fewer artifacts, finer edges and surfaces, as the highlighted regions shown in Figure 8. Quantitatively, our work provides a higher PSNR as **32.48** dB, **33.08** dB, **36.84** dB and SSIM as **0.8725**, **0.8859**, **0.9409** on **RA**, **LDP**, and **AI** respectively. Please check our supplemental video for the more intuitive comparison.

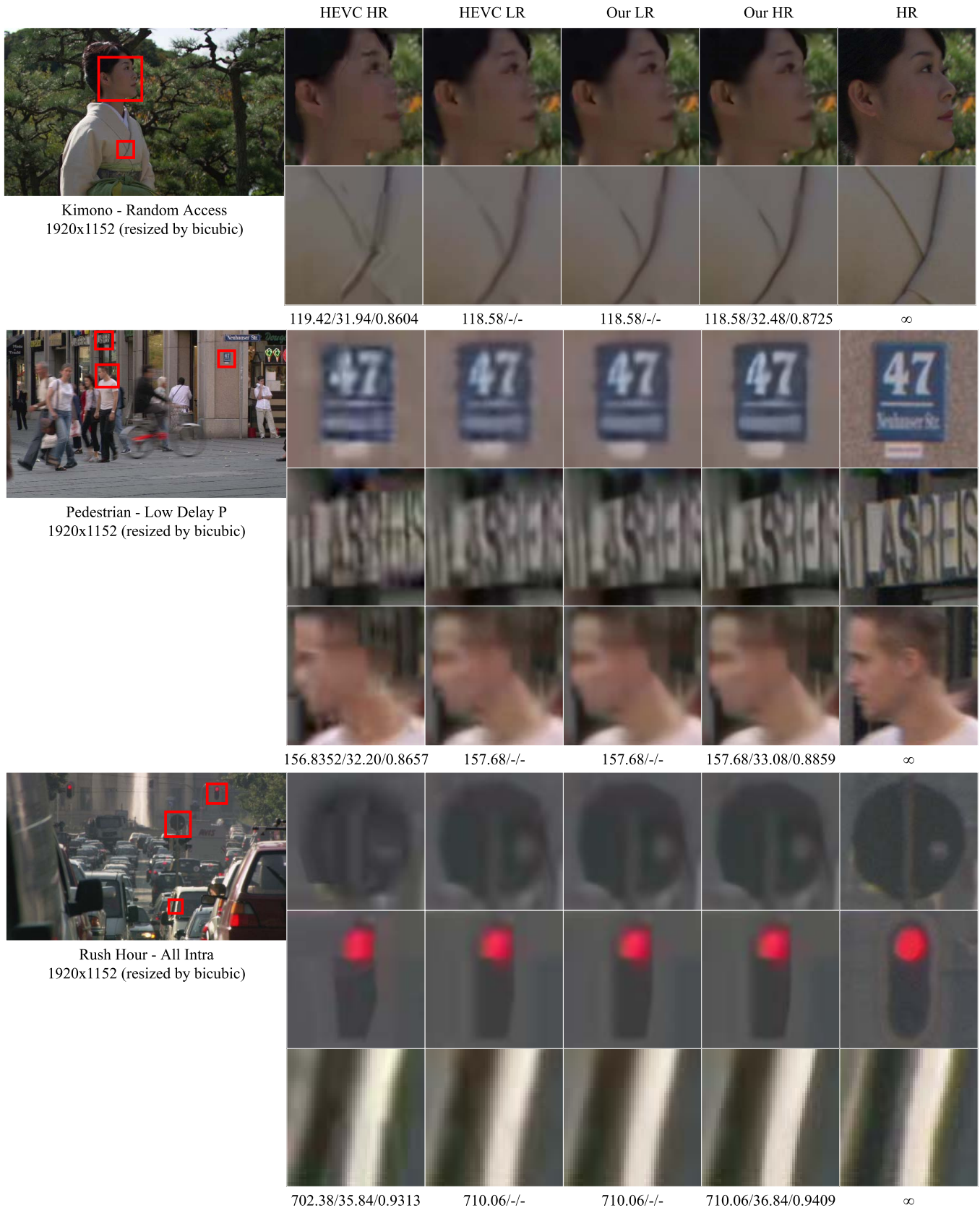

Fig. 8. Subjective comparison between our work and the standard HEVC in an approximate bit-rate condition. Additionally, we visualize our input (*HEVC LR*), our restored low-resolution (*Our LR*), and RR-DnCNN v2.0 (*Our HR*) in inference order. Our method outperforms in providing less video compression distortion as finer edges and surfaces in higher PSNR and SSIM (**bold** values). The measurement information is included for each method on each video as Bit-rate (kbps)/PSNR (dB)/SSIM. For easier concentration, we also highlight and zoom out several places corresponding to red rectangles.

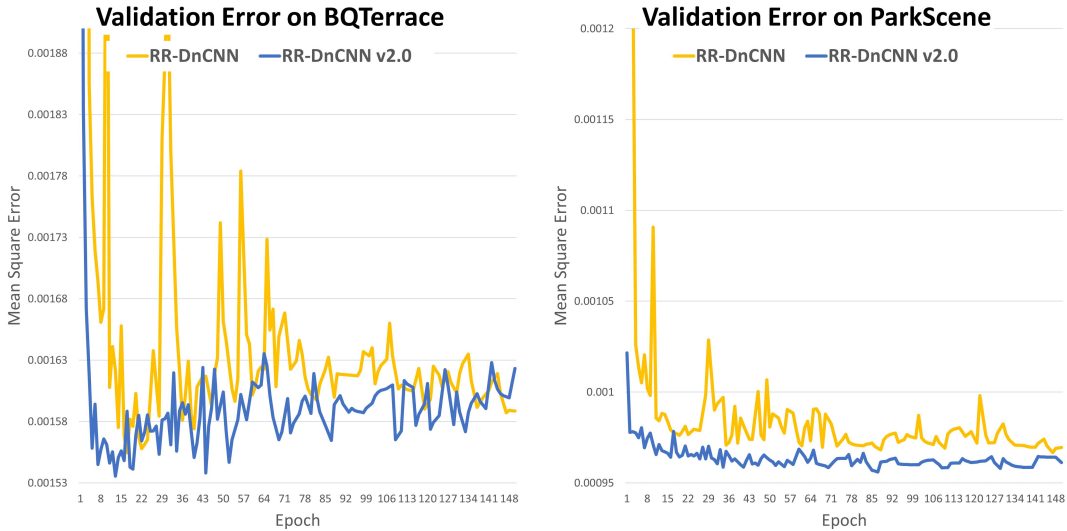

Fig. 9. Validation error on *BQTerrace* and *ParkScene*. The performance on *Park Scene* is converged stably epoch by epoch; meanwhile, *BQTerrace*'s validation error shows the over-fitting problem.

## IV. DISCUSSION

### A. Running on a Smaller Resolution Results in an Unstable Performance

Our scheme down-samples the spatial dimensions of a video to reduce the compression complexity; afterward, the video is restored, up-sampled, and reconstructed at the decoder's end. However, down-sampling a smaller-size video causes losing more useful information on the LR video for SR. It is even worse being compressed. Therefore, the larger-size video is, the more stably and effectively this scheme works. As shown in Tables V, VI, our method on sequences from class A,B, UHD [20] outperforms HEVC stably, especially on UHD. However, the results of the sequences from class C are unstable. For example, on the RA domain in Table V, our method achieves −6.10% and −10.72% on *BasketballDrill* and *RaceHorses*, respectively. However, in contrast to *BQMall* and *PartyScene*, the performance is dramatically worse than the standard HEVC as 1.02%, 18.03% BD-rate.

### B. Uncompensable Sequences

In this work, we utilize uncompressed videos to avoid unintended distortion and artifacts. Hence, our experiments are reliable. However, the number of uncompressed videos for training is limited; therefore, the overfitting problem may occur on some videos. For example, the validation errors on most of the test sequences converge stably while training, such as on *ParkScene*. Meanwhile, the validation error on *BQTerrace* becomes larger after several epochs and unstable. Although our RR-DnCNN v2.0 can improve the performance on *BQTerrace* compared to its old version, overfitting problem still occurs, as shown in Figure 9. In this SISR-based work, we successfully outperform the standard HEVC on AI as **-0.32%** BD-rate compared to our previous work [18], as shown in Table VII. Even so, the missing information gap on RA and LDP is currently not sufficiently compensated. Our future work will leverage neighboring frames. Besides, in industry, we can utilize more large-scale videos with various scenes for training, as long as the video quality is good enough.

## V. CONCLUSION

We improve our previous work [18] to have an end-to-end restoration-reconstruction u-shaped deep neural network (RR-DnCNN v2.0) with the degradation-aware technique. As a result, RR-DnCNN v2.0 outperforms its previous version [18] in learning capability as well as performance quantitatively and qualitatively, as shown in Table III and Figure 6. Furthermore, in compression proficiency BD-rate anchored by the standard HEVC, this work quantitatively outperforms the previous works [3], [18] as **−6.14%**, **−7.43%** on **RA**, **LDP** respectively, as shown in Table V, and **−10.67%** on AI compared to the works [2], [4], [18], as shown in Table VI. Negative BD-rates also prove our outperformance compared to the standard HEVC in overall. Furthermore, RR-DnCNN v2.0 can attain **17.02%** BD-rate reduction on the UHD sequences for AI. Outperforming [1] on three main configurations RA, LDP, and AI in all comparisons proves our proficiency in dealing with various degradation brought from video compression. Regarding our future work, we tend to leverage neighboring frames as MISR-based instead of this SISR-based work.

### TABLE VII
BD-RATE (%) ON THE VIDEO *BQTerrace* $1920 \times 1072$, ANCHORED BY THE STANDARD HEVC. ALTHOUGH THIS WORK OUTPERFORMS THE STANDARD HEVC ON AI, THE MISSING INFORMATION ON RA, AND LDP IS STILL NOT SUFFICIENTLY COMPENSATED

|     | RR-DnCNN [20] | RR-DnCNN v2.0 |
| --- | --- | --- |
| RA  | 25.44 | 24.85 |
| LDP | 32.55 | 31.21 |
| AI  | 0.66  | -0.32 |

---

[1] As [35] concerned, we clarify that there are many existing network architectures/designs combined with our degradation-aware technique which can probably achieve the same performance, especially the proposals in the super-resolution field.

TABLE VIII

Our Performance on Restoration for Low-Resolution Compared to HEVC and RR-DnCNN [18] on RA, LDP, and AI in PSNR (dB). This Work Outperforms the Previous Works on All Sequences in Class B and Additional B 1920 × 1072 (Cropped). Bold Value Denotes the Best Performance on Average

| Config | Sequence | HEVC | [20] | Ours |
|---|---|---|---|---|
| RA | BlueSky | 33.72 | 33.88 | 34.23 |
| | Pedestrian | 33.92 | 34.08 | 34.17 |
| | RushHour | 34.85 | 34.93 | 35.01 |
| | BQTerrace | 31.76 | 32.02 | 32.12 |
| | BasketballDrive | 33.65 | 33.87 | 33.92 |
| | Cactus | 31.69 | 31.93 | 32.03 |
| | Kimono | 32.86 | 32.96 | 33.01 |
| | ParkScene | 31.43 | 31.53 | 31.62 |
| LDP | BlueSky | 32.75 | 32.81 | 33.05 |
| | Pedestrian | 33.44 | 33.68 | 33.74 |
| | RushHour | 34.80 | 34.85 | 34.94 |
| | BQTerrace | 30.50 | 30.81 | 30.89 |
| | BasketballDrive | 33.42 | 33.69 | 33.72 |
| | Cactus | 31.01 | 31.22 | 31.31 |
| | Kimono | 32.24 | 32.31 | 32.33 |
| | ParkScene | 30.25 | 30.33 | 30.37 |
| AI | BlueSky | 33.74 | 34.07 | 34.37 |
| | Pedestrian | 35.35 | 35.65 | 35.80 |
| | RushHour | 37.08 | 37.21 | 37.37 |
| | BQTerrace | 30.73 | 31.05 | 31.13 |
| | BasketballDrive | 35.49 | 35.74 | 35.81 |
| | Cactus | 31.96 | 32.22 | 32.33 |
| | Kimono | 34.14 | 34.27 | 34.32 |
| | ParkScene | 31.33 | 31.46 | 31.53 |
| | Average | 33.01 | 33.19 | **33.30** |

## Appendix A
### Additional Comparisons Between RR-DnCNN [18] and RR-DnCNN v2.0

#### A. On Restoration for Low-Resolution

We show our performance on restoration at low-resolution compared to the previous work [18] and standard codec HEVC (HM 16.20) on three configuration RA, LDP, and AI in PSNR. As a result, our RR-DnCNN v2.0 outperforms on all sequences with the best average PSNR as **33.30** dB, as shown in Table VIII.

#### B. On Computational Complexity

We conduct the comparison on network architecture's complexity between RR-DnCNN [18] and this work in the number of parameters and multiply-add operations (MACs), as shown in Table IX. The increased number of parameters from [18] to this work is massively from deconvolutional modules in up-sampling skip connections.

## Appendix B
### When Will Our Model Well-Trained on QP = 37 Be Worse Than the Standard HEVC

We found that the QP = 37 produces the degradation not only rich enough to cover all QPs in the low bit-rate range QPs = {32, 37, 42, 47} but also not too distorted to make compensation possible. As a result, our R-D curve outperforms the standard HEVC's one at low bit-rates; however, our

TABLE IX

Computational Complexity Compared to Our Previous Work RR-DnCNN [18]. $M$: Million, $G$: Giga

| | Input size | Restoration length | Reconstruction length | Skip connections | Params (M) | MACs (G) |
|---|---|---|---|---|---|---|
| [20] | 640x360 | 10 | 15 | No | 0.82 | 520.97 |
| This work | 640x360 | 10 | 10 | Yes | 1.78 | 921.58 |

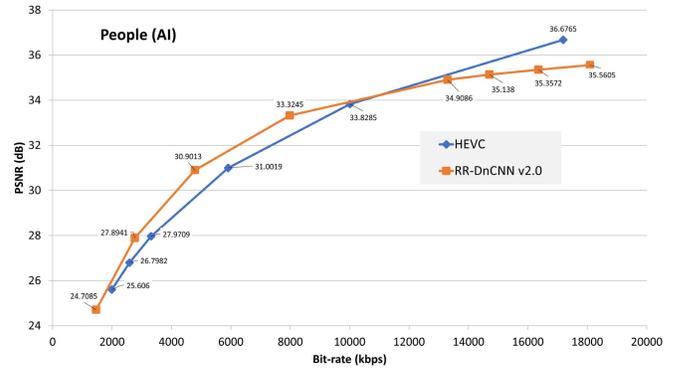

Fig. 10. We continuously check how our model trained on QP=37 performs on smaller QPs (QPs = {24, 25, 26, 27, 32, 37, 42, 47}) compared HEVC (QPs = {32, 37, 42, 47, 49, 51}) in RD-curves on the sequence *PeopleOnTheStreet* 1920 × 1152 (resized by bicubic) for all-intra. Our performance is reduced slower on higher bit-rate.

performance is reduced gradually on smaller QPs, as shown in Figure 10.

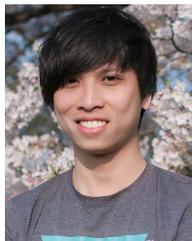

**Man M. Ho** received the B.S. degree (Hons.) in computer science from Vietnam National University, the University of Information Technology, Vietnam, in 2017, and the M.Eng. degree in applied informatics from Hosei University, Tokyo, Japan, in 2020. After graduation, he worked at EyeQ Tech, Vietnam, as a Machine Learning Engineer, from 2017 to 2018 and was recognized as a Key Contributor. He is currently a Research Student loving taking/editing/retouching photos. He is also a Graduate Student and a member of the Intelligent Media Processing Laboratory (IMPLab), Hosei University. His research interests include deep learning, computer vision, graphics, and photography.

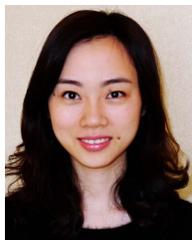

**Jinjia Zhou** (Member, IEEE) received the B.E. degree from Shanghai Jiao Tong University, China, in 2007, and the M.E. and Ph.D. degrees from Waseda University, Japan, in 2010 and 2013, respectively.

From 2013 to 2016, she was a Junior Researcher with Waseda University, Fukuoka, Japan. She is currently an Associate Professor and a Co-Director of the English-based graduate program with Hosei University. She is also a Senior Visiting Scholar with the State Key Laboratory of ASIC and System, Fudan University, China. Her research interests include algorithms and VLSI architectures for multimedia signal processing, including video coding, compressive sensing, computer vision, and VLSI design. She received the fellowship of the Japan Society for the Promotion of Science from 2010 to 2013. She received the Hibikino Best Thesis Award in 2011. She was a co-recipient of ISSCC 2016 Takuo Sugano Award for Outstanding Far-East Paper, the Best Student Paper Award of VLSI Circuits Symposium 2010, and the design contest award of ACM ISLPED 2010. She participated in the design of the world first 8K UHDTV video decoder chip, which was granted the 2012 Semiconductor of the Year Award of Japan. She has been a JST PRESTO Researcher since 2017.

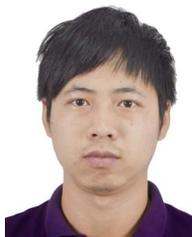

**Gang He** received the B.S. degree in electronic engineering from Xi'an Jiaotong University, Xi'an, China, in 2008, and the M.S. and Ph.D. degree in system LSI from Waseda University, Kitakyushu, Japan, in 2011 and 2014, respectively.

He currently works with the State Key Laboratory of Integrated Services Networks, Xidian University. He has authored or coauthored more than 20 papers in international journals and conferences. His current research interests include video coding algorithm and its VLSI architecture, image processing, and machine learning.